\documentclass{iaus}
\usepackage{graphicx}

\title[The ASTEP South project] 
{ASTEP South: \\ An Antarctic Search for Transiting Planets 
around the celestial South pole}

\author[N. Crouzet et al.]   
{N. Crouzet$^1$,
K. Agabi$^2$, A. Blazit$^2$, S. Bonhomme$^1$,\\Y. Fante\"\i -Caujolle$^2$, F. Fressin$^1$, T. Guillot$^1$, F.-X. Schmider$^2$,\\F. Valbousquet$^3$, E. Bondoux$^4$, Z. Challita$^4$, L. Abe$^2$, J.-B. Daban$^2$,\\C. Gouvret$^2$, and the ASTEP team}

\affiliation{$^1$Observatoire de la C\^ote d'Azur, Laboratoire Cassiop\'ee (UMR OCA-CNRS 6202), B.P. 4229, F-06304 Nice Cedex 4, France \\[\affilskip]
$^2$Observatoire de la C\^ote d'Azur, Laboratoire Fizeau (UMR OCA-CNRS 6525), Parc Valrose, F-06108 Nice Cedex 2, France \\[\affilskip]
$^3$Optique et Vision, 6 bis avenue de l'Est\'erel, BP 69, 06162 Juan-Les-Pins, France\\[\affilskip]
$^4$Concordia Station, Dome C, Antarctica \\[\affilskip]
email: {\tt Nicolas.Crouzet@oca.eu}}

\pubyear{2008}
\volume{253}  
\pagerange{}
\jname{Transiting Planets}
\editors{A.C. Editor, B.D. Editor \& C.E. Editor, eds.}

\begin{document}

\maketitle

\begin{abstract}

ASTEP South is the first phase of the ASTEP project that aims to determine the quality of Dome C as a site for future photometric searches for transiting exoplanets and discover extrasolar planets from the Concordia base in Antarctica. ASTEP South consists of a front-illuminated 4k x 4k CCD camera, a 10 cm refractor, and a simple mount in a thermalized enclosure. A double-glass window is used to reduce temperature variations and its accompanying turbulence on the optical path. The telescope is fixed and observes a $4\,^{\circ}$ x $4\,^{\circ}$ field of view centered on the celestial South pole. With this design, A STEP South is very stable and observes with low and constant airmass, both being important issues for photometric precision. We present the project, we show that enough stars are present in our field of view to allow the detection of one to a few transiting giant planets, and that the photometric precision of the instrument should be a few mmag for stars brighter than magnitude 12 and better than 10 mmag for stars of magnitude 14 or less.

\end{abstract}

\firstsection 
\section{Introduction}

The ASTEP project (Antarctic Search for Transiting Extrasolar Planets) aims to  determine the quality of Dome C, Antarctica as a site for future photometric surveys and to detect transiting planets (\cite[Fressin et al. 2005]{}). The 3 month continuous night as well as a very dry atmosphere should yield to a great improvement of the photometric precision when compared to other sites. The goal of ASTEP South is to qualify the photometry from Dome C and to attempt to detect planets. First, the instrument setup is described. We then present the expected planet detection probability using a simple model. Finally, we use SimPhot to simulate the ASTEP South observations and a first data reduction process. The expected noise is then derived for stars in the ASTEP South field of view.

\section{The ASTEP South setup}

\begin{figure}[ht]
\centering
\begin{tabular}{ccc}
\resizebox{7.5cm}{!}{\includegraphics{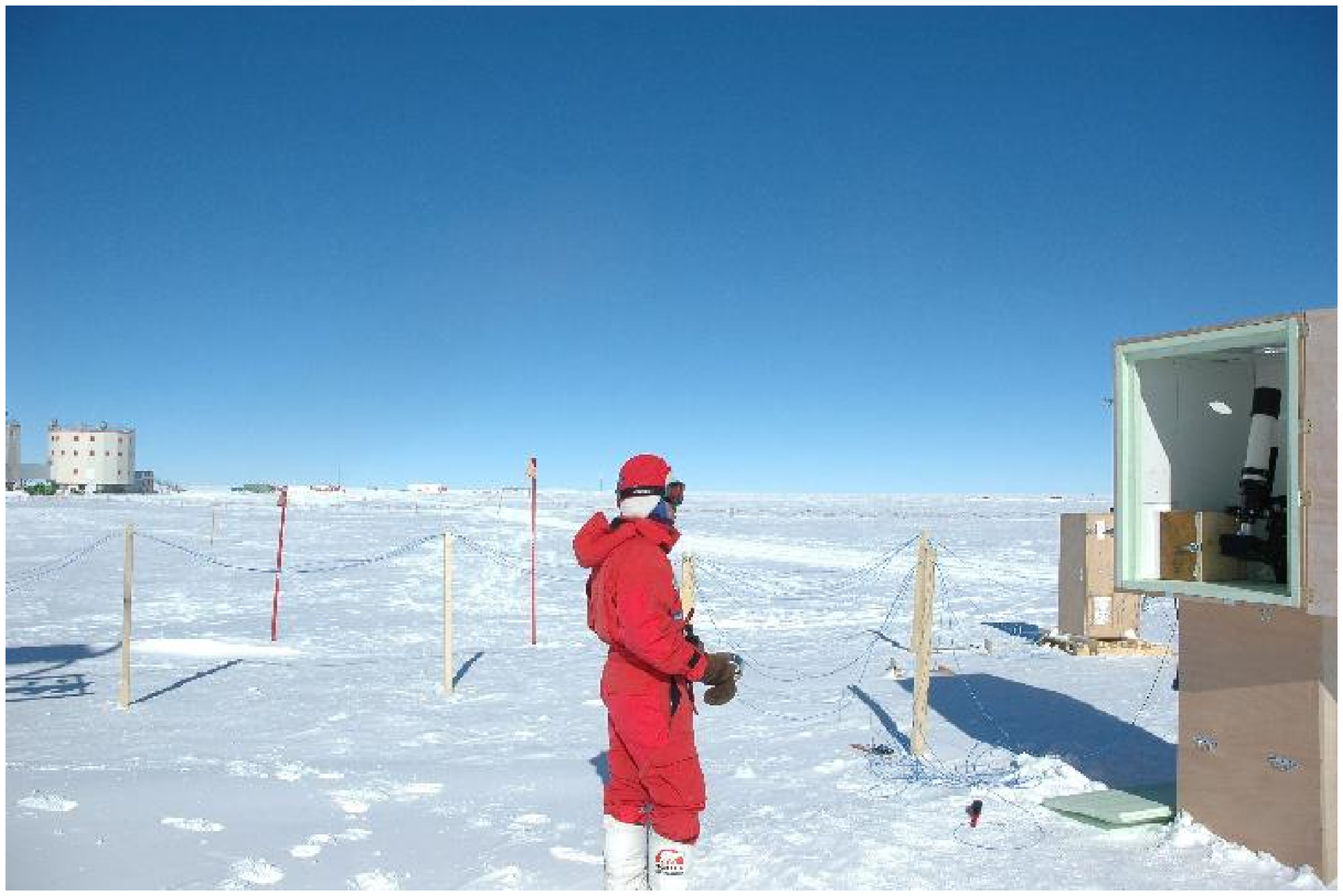}} & \hspace{0.5cm} &
\resizebox{5cm}{!}{\includegraphics{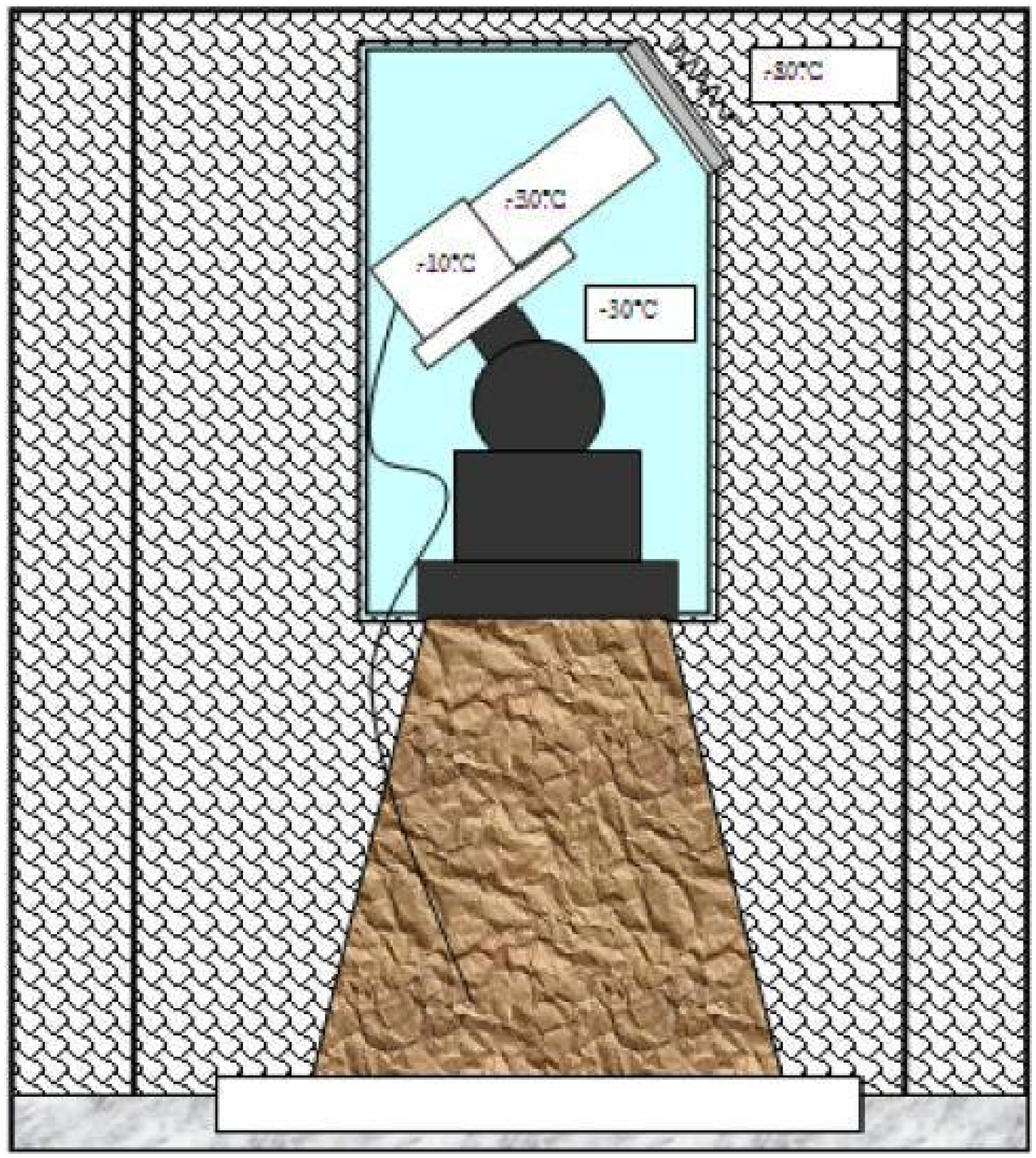}} \\
\begin{minipage}[t]{7cm}
\caption{ASTEP South at Dome C, Antarctica, January 2008.}
\label{fig:ASTEPSouthDomeC}
\end{minipage} & \hspace{0.5cm} &
\begin{minipage}[t]{5cm}
\caption{The ASTEP South design : a 10 cm refractor, a front-illuminated 4k x 4k CCD camera, and a simple mount in a thermalized enclosure. The instrument is completely fixed, avoiding jitter noise.}
\label{fig:setupASUD}
\end{minipage} 
\end{tabular}
\end{figure}

The instrument is composed of a 10 cm refractor, a front-illuminated 4k x 4k CCD camera (see \cite[Crouzet et al. 2007]{} for the choice of the camera), and a simple mount in a thermalized enclosure. A double glass window is used to reduce temperature variations and its accompanying turbulence on the optical path. The ASTEP South instrument is shown at Dome C in figure~\ref{fig:ASTEPSouthDomeC} and detailed in figure~\ref{fig:setupASUD}. In order to characterize the quality of Dome C for photometric observations, we have to avoid as much as possible instrumental noises and in particular jitter noise, leading to a new observation strategy: the instrument is completely fixed and points towards the celestial South pole continuously. The observed field of view is $3.88\,^{\circ}$ x $3.88\,^{\circ}$ centered on the celestial South pole and contains around 8000 stars up to Mv=15. This observation setup leads to (i) stars moving on the CCD and coming back to the same pixels every sideral day, and (ii) an increase of the PSF (Point Spread Function) size in one direction and a greater contamination, depending on the exposure time.\\
Test observations were made at the Calern site observing the celestial North Pole, in order to choose the exposure time and the PSF size. A 30 second exposure time and a 2 pixel PSF FWHM (Full Width Half maximum) lead to only 2 saturated stars and a limiting magnitude of 14. An analysis of the celestial South Pole field from the Guide Star Catalog with the same parameters leads to less than 10 \% of contaminated stars. Therefore we adopted these parameters. The instrument was set up at the Concordia base in January-February 2008. ASTEP South is now observing and delivering its first data.

\section{A simple model}
\begin{figure}[ht]
\centering
\resizebox{9cm}{!}{\includegraphics{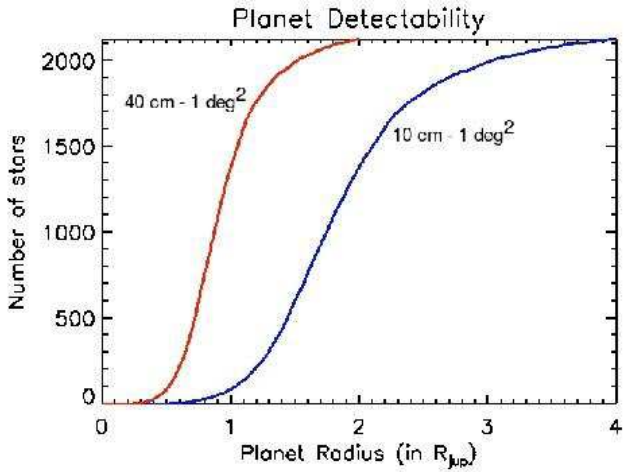}}
\begin{minipage}[c]{9cm}
\caption{Number of stars for which a planet equal to or larger than a given radius can be detected, for a $1\,^{\circ}$ x $1\,^{\circ}$ field of view around the South Pole. For comparison purpose, results for a 10 cm refractor and a 40 cm telescope are shown.}
\end{minipage}
\label{fig:count_stars}
\end{figure}

We perform an analytical model to evaluate the planet radius that can be detected with ASTEP South. Simulated field at the celestial South pole from the Besan\c con model are used. We consider that a planet is detected if the flux variation is at least 3 times the photon noise. From the star radius, we obtain the minimum planet radius that can be detected. As shown in figure~\ref{fig:count_stars}, in a $1\,^{\circ}$ x $1\,^{\circ}$ field of view, a planet with a radius of 1.5 $R_{jup}$ can be detected for roughly 500 stars. For our $3.88\,^{\circ}$ x $3.88\,^{\circ}$, this number grows to 7000. For comparison, the number of F, G, K dwarfs per transiting giant planet is estimated to be about 1100 to 1 (\cite[Fressin et al. 2007]{}). This implies that enough stars are present to allow the detection of one to a few transiting giant planets.

\section{A numerical study}
\begin{figure}[ht]
\centering
\begin{minipage}[c]{7cm}
\resizebox{7cm}{!}{\includegraphics{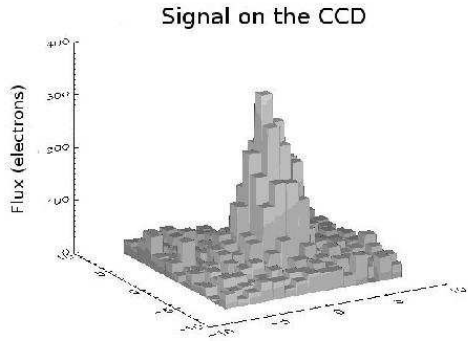}}
\end{minipage}
\begin{minipage}[c]{6cm}
\caption{Simulated magnitude 15 star for a 40 cm telescope and a 30 second exposure time. After passing through the atmos\-phere and the instrument, the si\-mulated signal is processed with aperture photo\-metry and compared with a reference star.}
\end{minipage}
\label{fig:pixels}
\end{figure}

\begin{figure}[ht]
\centering
\resizebox{9cm}{!}{\includegraphics{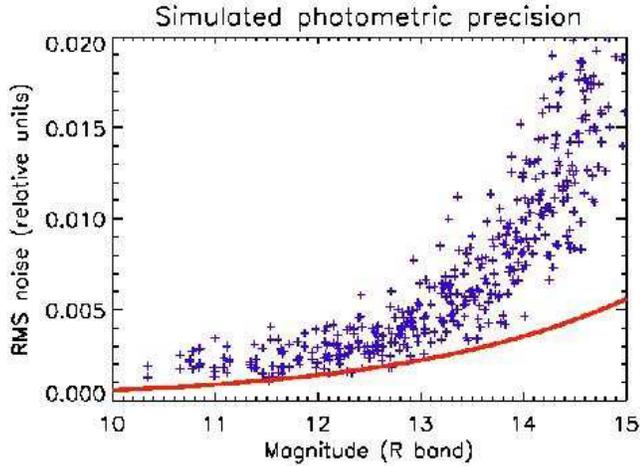}}
\begin{minipage}[c]{9cm}
\caption{Simulated rms noise against magnitude using SimPhot with the ASTEP South parameters, for 450 target stars in the $1\,^{\circ}$ x $1\,^{\circ}$ central part of the South Pole field. The continuous line shows the photon noise limit.}
\end{minipage}
\label{fig:rms-mag}
\end{figure}

In order to further test the ASTEP South sensitivity, we use SimPhot, a photometric simulator that aims to reproduce each step of a survey from the observations to the final lightcurves. The starting point is the flux from target and background stars. Atmospheric elements such a seeing variations and sky background are added. A CCD transmission matrix is then simulated as well as instrumental noise such as PSF variations on the CCD. Simulated images are then reduced using aperture photometry and comparison with a reference star. Lightcurves and photometric precision for each target star are then obtained. Figure~\ref{fig:pixels} shows a simulated image for a magnitude 15 star for a 40 cm telescope and a 30 second exposure time. For a more detailed analysis of some of the noise sources included in SimPhot see \cite[Crouzet et al. 2007]{}.\\
For each target star in the $1\,^{\circ}$ x $1\,^{\circ}$ central part of the South Pole field, we simulate 2 hours of observation using the ASTEP South parameters. The flux in the photometric aperture is smoothed over 20 minutes. A comparison star of the same magnitude was used to correct for seeing variations. The effect of different PSF for both stars is included. The rms noise obtained with these simulations is plotted against magnitude in figure~\ref{fig:rms-mag}. This shows that a few mmag precision should be reached for stars brighter than magnitude 12 with a noise level close to the photon noise, and that the noise level is better than 10 mmag for stars of magnitude 14 or less.

\section{Conclusion}

ASTEP South is the first transit survey from Dome C, Antarctica. The instrument inside its thermalized box is currently observing towards the celestial South pole. A simple analytical model shew that the observed field of view contains enough stars to enable transit detections with our 10 cm refractor. Simulations made with SimPhot show a rms noise close to the photon noise for stars of magnitude 12 or less, and a noise level better than 10 mmag for stars of magnitude 14 or less. Thus, this first campaign should allow us to test this new observing method, to qualify Dome C for photometric observations and possibly to detect planets.

\end{document}